# The prototypical organic-oxide interface: intra-molecular resolution of sexiphenyl on In$_2$O$_3$(111)

Margareta Wagner*[a], Jakob Hofinger[a], Martin Setvín[a], Lynn A. Boatner[b], Michael Schmid[a], and Ulrike Diebold[a]

[a] Institute of Applied Physics, TU Wien, Wiedner Hauptstraße 8-10/134, 1040 Vienna, Austria
[b] Materials Science and Technology Division, Oak Ridge National Laboratory, Oak Ridge, Tennessee 37831, USA

* Corresponding author: wagner@iap.tuwien.ac.at

*Abstract*

The performance of an organic-semiconductor device is critically determined by the geometric alignment, orientation, and ordering of the organic molecules. While an organic multilayer eventually adopts the crystal structure of the organic material, the alignment and configuration at the interface with the substrate/electrode material is essential for charge injection into the organic layer. This work focuses on the prototypical organic semiconductor *para*-sexiphenyl (6P) adsorbed on In$_2$O$_3$(111), the thermodynamically most stable surface of the material that the most common transparent conducting oxide, indium tin oxide (ITO) is based on. The onset of nucleation and formation of the first monolayer are followed with atomically-resolved scanning tunneling microscopy (STM) and non-contact atomic force microscopy (nc-AFM). Annealing to 200°C provides sufficient thermal energy for the molecules to orient themselves along the high-symmetry directions of the surface, leading to a single adsorption site. The AFM data suggests a twisted adsorption geometry. With increasing coverage, the 6P molecules first form a loose network with poor long-range order. Eventually the molecules re-orient and form an ordered monolayer. This first monolayer has a densely packed, well-ordered (2×1) structure with one 6P per In$_2$O$_3$(111) substrate unit cell, *i.e.,* a molecular density of 5.64 × 10$^{13}$ cm$^{-2}$.

Keywords: indium oxide, sexiphenyl, organic molecules, atomic force microscopy, scanning tunneling microscopy, adsorption site, monolayer.

**Introduction**

Transparent conducting oxides (TCOs) such as ZnO, CdO, $SnO_2$, $Ga_2O_3$, and $In_2O_3$ uniquely combine transparency in the visible range of light with electric conductivity as wide-band-gap semiconductors (fundamental band gap > 2 eV).[1] In the case of $In_2O_3$ (band gap of ~2.9 eV, see Ref. 2 for an overview) the intrinsic free-electron concentration at room temperature (RT) is already $10^{19}$ cm$^{-3}$,[3] and can be further enhanced by doping. One of the most widely used dopants is tin (indium tin oxide, ITO), which makes the material almost metal-like in terms of its conductivity while maintaining its optical transparency. ITO is extensively used as electrode material in light emitting diodes (LEDs), solar cells, and liquid-crystal displays. Beyond its leading role as a transparent electrode material, $In_2O_3$ and ITO are also used, *e.g.,* as sensor material to detect gases.[4,5] In this work we use undoped $In_2O_3$ single crystals as a model system; in previous work we have established that the surfaces of ITO(111) and $In_2O_3$(111) are very similar.[6,7]

In the late 1990s, one of the first blue organic light emitting diodes (OLEDs) was fabricated from the organic semiconductor *para*-sexiphenyl with ITO as the anode material.[8,9] Today 6P is known as a model molecule for use in organic devices with a high quantum yield and chemical stability adequate for device technology. The pure powder material is suitable to be handled in an ultrahigh vacuum (UHV) environment, and can be sublimed at ~210 °C. In its crystalline form at least three bulk structures have been reported for different temperature ranges.[10,11] The most commonly observed β-phase (formed at RT) is monoclinic with the space group $P2_1/a$, two molecules per unit cell and lattice parameters of a = 0.8091 nm, b = 0.5568 nm, c = 2.6241 nm, and β = 98.17°. The 6P molecules have their long axis parallel to each other but are inclined by 17° to the c-axis of the unit cell.[11] In a plane perpendicular to the molecular long axis, *i.e.*, roughly 6P(001) or the top-view of an upright standing layer, a herringbone arrangement is formed of alternatingly tilted molecular planes in adjacent 6P(20-3) lattice planes. The phenyl rings of the β-phase are considered to be planar, i.e., not twisted. The low-temperature α-phase (110 K) was argued to be very similar to the β-phase, but featuring twisted molecules.[10,11] Recently, a γ-phase was observed at a growth temperature of 160 °C with small variations in the lattice parameters compared to the β-phase and with molecules parallel to the c-axis.[11]

The onset of crystalline growth of 6P supported on surfaces, as well as its geometric and electronic properties have been studied extensively on numerous noble metals, on surface oxides including the (2×1)O reconstruction of Cu(110) and Ni(110), as well as on a few bulk insulators[12,13] and oxides, mainly rutile TiO$_2$(110).[14-16] Oxides are often (but not always correctly) assumed to interact weakly (as compared to metals) with such organic compounds, which should—in a first approximation—result in densely packed structures of tilted and/or upright standing molecules, where the π-overlap between the molecules is maximized and the surface is only slightly wetted. However, even on TiO$_2$(110) a wetting layer of flat-lying but tilted molecules was observed, where the molecules were aligned in [001] direction in a (very likely) commensurate formation.[17] Successive growth strongly depends on the growth temperature, resulting in a multilayer of flat-lying molecules at room temperature with the 6P(20-3) plane parallel to the TiO$_2$ substrate surface. Deposition at elevated temperatures results in almost upright standing molecules (*i.e.,* where the 6P(001) plane is parallel to the substrate surface), without restructuring of the first layer.

In this work we discuss the adsorption of the single 6P molecules on In$_2$O$_3$(111) and the formation of the first, densely packed (2×1) monolayer. The In$_2$O$_3$(111) surface has a 3-fold symmetry and an in-plane lattice parameter of 1.43 nm. The unit cell contains 40 atoms in an O-In-O tri-layer arrangement, *i.e.,* 16 indium atoms are surrounded by 2 × 12 oxygen atoms. In the bulk In is only present with a 6-fold coordination, In(6c). At the surface, however, four indium atoms are In(6c), forming a three-pointed star in the unit cell (imaged as dark triangle in STM); the remaining 12 atoms are In(5c). Of a total of 24 O atoms within the surface unit cell, the 12 atoms above and 12 below the In plane are present as O(3c) and O(4c), respectively; the surface is thus oxygen terminated. In empty-states STM images, the region containing the In(6c) atoms appears as a depression due to a low local density of states.[7] Sexiphenyl (C$_{36}$H$_{26}$, hexaphenyl, 6P) is a rod-like, organic molecule consisting of 6 phenyl rings in *para* configuration. The Van der Waals size of 6P measures 2.72 × 0.67 nm$^2$,[10] thus its length is almost twice the lattice parameter of In$_2$O$_3$(111).

Differing from the growth on other metal and oxide surfaces, the ordering of the 6P molecules on In$_2$O$_3$(111) is poor at RT, *i.e.,* the molecules do not orient themselves with respect to a high-symmetry direction of the surface. To overcome the diffusion barrier of the molecules and also to clean the surface from OH groups, a post-annealing step is required, resulting in oriented molecules that are present in three equivalent orientations (due to the 3-fold symmetry of In$_2$O$_3$(111)), with all of the molecules featuring the same adsorption site. Thus, starting with single, randomly distributed but oriented molecules, the formation of a poorly ordered, dilute layer is observed. Upon further deposition, the dilute layer reorganizes into a well-defined, densely packed (2×1) monolayer. Successive growth leads to a (1×1) structure with respect to the In$_2$O$_3$(111) surface.

The molecules of the first layer 6P on In$_2$O$_3$(111) are lying flat on the surface, although their adsorption and appearance in STM differs significantly from what has been observed on other substrates. While the flat-lying 6P molecule is usually imaged as a stiff, rod-like object (occasionally with internal zigzag-like protrusions due to twisting of the phenyl rings[18,19]) its shape on the In$_2$O$_3$(111) surface resembles a slightly asymmetric "W". Also in the constant-height AFM data obtained with a CO functionalized tip the 6P molecule differs in appearance from what one would expect for a flat-lying molecule. AFM clearly shows that 5 of the 6 phenyl rings lie rather flat, and provides the adsorption site of the molecule.

**Experimental**

The STM experiments were carried out in a two-chamber UHV system equipped with an Omicron LT-STM cooled with liquid nitrogen and operated at ~80 K. Electrochemically etched W tips were initially prepared by sputtering with Ar followed by scanning on Au(110). Voltage pulses were applied until clear atomic resolution and sharp step edges were obtained as well as a metallic signature in the scanning tunneling spectroscopy signal. The tip was treated similarly on In$_2$O$_3$(111) as well and was frequently refreshed on gold. Additional experiments were conducted in a commercial Omicron qPlus LT-STM at 5 K, using tuning-fork sensors[20] with a separate wire for the tunneling current[21] ($k$ = 3,750 N/m, $f_R$ = 47,500 Hz, $Q \approx$ 10,000) and a differential preamplifier.[22] Electrochemically-etched W tips (20 μm wire size) were glued to the tuning fork. These were cleaned in UHV by self-sputtering and field emission,[23] and on a Cu(100) surface by voltage pulses until a metallic behavior and a frequency shift $\Delta f$ between 0 and -3 Hz was reached (set point 30 pA, +1 V). For functionalization of the tip, small amounts of CO were dosed into the STM and individual CO molecules picked up from the $In_2O_3$ surface. Imaging 6P with a CO tip was tested on Cu(100), (see the Supporting Information). The data presented here are constant-current STM images probing the empty states, and constant-height non-contact AFM measurements probing short-range forces. The imaging contrast due to short-range forces is composed of attractive (dark, strongly negative frequency shift) and repulsive (bright, less negative frequency shift) interactions with the AFM tip. During the AFM measurements, a small positive bias voltage was applied, and the mean tunneling current $<I_T>$ was simultaneously recorded. On $In_2O_3$(111), 0.4-5 mV was applied for the metallic tips, and 50-100 mV was used for CO tips.

Single crystals of indium oxide,[24] cut and polished along the thermodynamically most stable (111) plane, are used here as substrates. The $In_2O_3$(111) single crystals were mounted on sample plates made of Ta. The crystal surface was cleaned by several cycles of sputtering and annealing. The sputtering was carried out in normal incidence with a focused ion gun (SPECS) scanning across the crystal surface only (1 keV $Ar^+$, ~1.6 µA sample current). The annealing at ~450 °C was alternatingly performed in UHV (reducing conditions) or in oxidizing conditions, by backfilling the chamber with ~$6\times10^{-7}$ mbar $O_2$; this pressure was kept until the crystal was cooled to 150 °C. The final annealing was carried out in $O_2$ in order to obtain a stoichiometric, relaxed bulk-terminated surface as described in Ref. 7.

Sexiphenyl was deposited *via* thermal sublimation from powder (Tokyo Chemical Industry Co., Ltd.) using a water-cooled four-pocket evaporator (Omnivac) as well as a home-built single evaporator; in both cases, the crucible was heated indirectly by a filament. The deposition rate was monitored with a quartz crystal micro balance (QCM) positioned directly in front of the sample, and the temperature of the evaporator (~210/230 °C measured at the bottom/top of the crucible of the commercial/home-built evaporator) was set to achieve a deposition rate of ~1-2 Å/min (density of 6P in the β-phase: 1.3 g $cm^{-3}$). During deposition, the sample was kept at room temperature and was afterwards annealed at 200 °C for 3 min. This final annealing was used because deposition of 6P at RT results in disordered and randomly oriented molecules. Possibly water co-evaporation contributes to the disordered character of the layer (see Supporting Information). The post annealing allows the molecules of the first layer to diffuse, reorient, and align themselves on the surface; additionally, hydroxyl groups (originating from dissociated water from the residual gas and the evaporator) desorb from the surface. Depositing molecules directly at a substrate temperature of 200°C leads to the same results as deposition at RT followed by post-annealing. At this temperature, molecules of successive layers already desorb; this allows for the easy preparation of the densely packed layer by flashing off the multilayer. Higher coverages were thus annealed at 120 °C only.

**Results**

The growth of the first layer of sexiphenyl on $In_2O_3$(111) can be divided into two regimes in terms of the orientation and assembly of the molecules.

At low, sub-monolayer coverages, as shown in Figures 1(a, b), single molecules are randomly distributed across the surface, with no apparent preference for step edges or defects. However, each individual molecule is clearly oriented along one of three equivalent crystallographic <1-10> directions. The adsorption site of the molecules is found to be in-between the rows formed by the dark triangles of the clean $In_2O_3$(111) surface.

In addition to the STM measurements, the single molecules were investigated with nc-AFM. Figure 2 shows a comparison of different imaging modes; here a metallic tip was originally prepared but picked up a 6P molecule or a fragment thereof during scanning. The shape of the 6P molecules in constant-current and constant-height STM (panels (a, b)) resembles an asymmetric "W", where the asymmetry in shape is identical for all molecules of the same orientation. In the constant-height frequency-shift image, panel (c), the molecules look more rod-like in general; but their internal structure again shows an asymmetry along the 6P long axis. One end of the molecule, *i.e.,* the last phenyl ring, is imaged as a distinct, bright semi-circle (red arrows in Figure 2(c)) due to repulsive interactions with the AFM tip, while the other end of the molecule blends into the surface. Since all of the molecules appear to be identical in STM and AFM, despite their different orientation, a single adsorption site and geometry can be deduced.

To obtain further information about the other phenyl rings of the molecule and the exact adsorption site on the $In_2O_3$(111) surface, CO was co-dosed into the STM/AFM at 5 K and picked up by the tip, which allowed us to image the "chemical structure/backbone" of the molecules in constant-height measurements. The result is depicted in Figure 3(a). Although the molecule looks very different from what one would expect for a planar and flat molecule (see 6P on Cu(100) in the Supporting Information), the individual phenyl rings are easily recognized. For better visibility a high-pass filter was applied to the images in panel (b). The bright semi-circle from Figure 2(c), labeled "6", is readily identified as a flat-lying phenyl ring, as are rings "2" and "3" and "5". Ring "1" is only faintly visible, but a gentle push with the tip during image acquisition revealed its nature as an intact phenyl ring, see Figure 4. The configuration of ring "4" remains unclear. The strong interaction between the tip and ring "4" could be due to an asymmetrically adsorbed CO molecule at the tip, or a twisted configuration of the molecule. The tunneling current $<I_T>$, recorded simultaneously with the frequency shift $\Delta f$ of the constant-height AFM measurement is displayed in Figure 3(c), and for a slightly larger region in panel (e). Contrast enhancement reveals features of the $In_2O_3(111)$ surface, *i.e.*, the dark triangles located at the In(6c), which dominate the STM contrast on the uncovered surface; see also Figures 1 and 2. The superposition of the frequency shift and tunneling current signals with the (relaxed, bulk terminated) atomic model of the bare $In_2O_3(111)$ surface allows postulating a tentative adsorption geometry of a (simplified) flat-lying and planar sexiphenyl on $In_2O_3(111)$, see Figure 3(d). The error in position due to the alignment of Figure 3(e) with the lattice is about 0.1 nm.

Increasing the coverage does not influence the adsorption until the whole surface is loosely covered. The molecules arrange into a so-called "dilute layer" where every molecule occupies the same adsorption site, while the whole assembly lacks long-range order due to the three possible and equivalent <1-10>-type orientations. Domains remain small and consist of (at most) 5 to 6 molecules, see Figure 1(c). Approaching the density of one 6P per substrate unit cell, however, results in a reorientation of the molecules away from <1-10> to form a new structure. The onset of this nucleation is depicted in Figure 5(a), where densely packed stripes (marked by yellow arrows) are surrounded by molecules that have already re-oriented, but are not yet part of the new structure (black arrow). A surface fully covered by this new structure is shown in Figure 5(b). Three domains are observed, each consisting of bright, "dashed" stripes running along one of the three high symmetry <1-10> directions of the $In_2O_3(111)$ surface, forming a (2×1) superstructure. Figure 6(a) shows a large domain of the new structure and its border (white arrows), where some molecules oriented differently. The (2×1) unit cell is compared to the $In_2O_3(111)$ cell in Figure 6(b).

The appearance of the (2×1) superstructure depends on the status of the STM tip. The most common contrast of a metallic STM tip, obtained after preparing the tip on a gold surface, is depicted in Figures 5(b, top) and 6(c); here the individual molecules in the darker stripes are hardly recognizable. However, poking the tip into a surface with high 6P density, and presumably by picking up a molecule or fragment, always leads to similar contrasts on the (2×1) structure; here the individual molecules are easily distinguished (Figures 6(d, e)). On close inspection, both, the bright "dashed" stripes and the dark spaces in-between consist of densely packed 6P molecules, enclosing the angle $\alpha = \beta = 30°$ to the stripe direction (indicated in Figure 6(e)). This means that the molecules are symmetrically arranged around the <1-10> directions of the surface, with their long axes aligned in <-211>, *i.e.*, the diagonal of the $In_2O_3$ unit cell. The (2×1) unit cell contains two molecules, one from the dark and one from the bright stripe. This arrangement is obvious from the high-resolution images with the functionalized tip, but is not so clear in the contrast obtained with metallic tips, Figure 6(c). The elongated protrusions forming the bright stripes of Figure 6(c) also seem to enclose a different, larger, angle to the stripe direction. Our analysis (see Supporting Information) shows that these protrusions in the contrast obtained with metallic tips are, in fact, not located exactly at the position of the molecules. This is also indicated by the (2×1) unit cell in panels (c-e), which is positioned with respect to the molecules of the dark stripes. In panel (c) the protrusions appear in-between the molecules of the bright stripes.

The "dilute layer" and the (2×1) monolayer are the only stable structures after annealing at ~200 °C. Depositing small amounts of additional 6P at room temperature on top of the (2×1) layer leads to a mostly disordered phase with some ordered patches but with an overall unstable surface during STM imaging. An ordered phase can be produced by gentle annealing at ~120 °C. In STM large protrusions forming a (1×1) pattern are observed, see Figure 5(c). The apparent height of these protrusions with respect to the surrounding surface (probably the densely packed layer) is 140-170 pm. A further analysis of this structure is beyond the scope of the current work.

**Discussion**

In the present work, we have used single crystals of pure $In_2O_3$ as a model system to investigate the early stages of 6P growth on the $In_2O_3$(111) surface. The growth process is followed using scanning tunneling microscopy (STM) and non-contact atomic force microcopy (nc-AFM) at cryogenic temperatures. The use of nc-AFM, operated at constant height with a CO-functionalized tip is an invaluable tool in discerning the chemical structure of organic molecules, not only on insulating substrates, but more recently also on metals and oxides.[25,26] Recent overviews have been published by Jarvis[27] and Jelínek.[28] The sharp contrasts obtained with a CO functionalized tip on organic molecules result from a combination of Pauli repulsion and the tilting of the flexible CO-tip due to electrostatic forces.[26] Thus, the impression of the geometric structure of a molecule can be distorted by the imaging mechanism itself; other effects can also contribute, e.g., dipole moments of the substrate or charge rearrangements within the molecule.[29]

After deposition at room temperature, single sexiphenyl molecules adsorb disordered on the $In_2O_3$(111) surface. To overcome the diffusion barrier and desorb the OH groups, a post-annealing step was performed. The temperature range of 150-200 °C provides enough thermal energy to align the 6P molecules on the surface without desorbing or destroying them. Additionally, hydroxyl groups due to dissociative water adsorption are removed from the $In_2O_3$(111) surface,[30] which could potentially also be a factor in the initially-observed random orientation of the molecules.

Two structures that were observed in the first monolayer differ in the molecular density, orientation and adsorption sites of the molecules, as well in their long-range order. At lower coverage, a structure called here the "dilute layer", is formed with very small domains, where all of the 6P adopt the same adsorption site and are oriented along the same, equivalent low-index <1-10> directions. The molecules are well-separated from each other, except for linear chains where their ends meet. Molecules of mixed orientations enclose 60°, and the end of one molecule can meet either the end or the side of the next molecule (see Figure 1 (b, c)). Overall, apart from steric hindrance, they are not influenced by the presence of neighboring molecules, which is manifested in the lack of domain formation and long-range order. The appearance of the individual molecule in constant-current STM resembles an asymmetric "W", *i.e.*, there is a significant zigzag in the shape of the molecule. This differs from the way 6P usually appears in STM,[18,19,31] and leads to the suggestions that the molecule does not adsorb in a completely flat configuration on $In_2O_3(111)$. Sexiphenyl is a structurally flexible molecule due to the single C-C bond linking the phenyl rings. This allows the molecule to bend out of plane, *e.g.* across step edges, while in-plane bending (without twisting) is limited by the sp2-hybridisation of the C-C bond and steric hindrance of the hydrogen atoms. However, 6P can twist, *i.e.*, individual phenyl rings can be rotated out of plane, usually to increase the π-π overlap between neighboring molecules, which is observed in the 6P crystal structure (α-phase), and, supported on surfaces, both in multilayers[19] and monolayers.[31]

While the configuration of the molecule remains unclear from the STM images, the constant-height AFM data obtained with a CO tip deliver more details. It should be noted that the following interpretation is based purely on experimental evaluation and could serve as a useful input for future theoretical modeling. Moreover, the error in positioning is about 0.1 nm. Similar to the STM images, the appearance of 6P in constant-height AFM differs from what has been seen previously for planar and flat, rod-like conjugated molecules on metals and insulating surfaces, even if the registry of the phenyl rings with the substrate changes along the molecule.[25,32,33] On $In_2O_3(111)$, the 6P molecule shows phenyl rings in a rather flat geometry (the hexagons are discernable) superimposed by pronounced variations in frequency shift along the molecular long axis, see Figure 3(a). The maximum (brightest contrast) is located at phenyl ring "4" (strongly interacting with the tip), followed by pairs of equal contrast, first by "3" and "6", then "2" and "5", and finally ring "1", which interact attractively/very weakly (dark). This up-and-down configuration can be explained by the surface structure and oxygen termination of $In_2O_3(111)$. The surface of $In_2O_3(111)$ consists of an O(4c)-In(5c, 6c)-O(3c) tri-layer, where the O(3c) are the topmost atoms, see Figure 3(d). The In(6c) atoms (blue in Figures 3(d)), are clustered in a three-pointed star and appear as dark triangles in the STM images. The surface termination is not homogeneous; it is either O(3c) or In(5c). 6P adsorbs in a position where it touches two In(6c) triangles (blue in Figure 3(d)). It covers only the O(3c) terminated regions next to the In(6c) (phenyl ring "3" and "6"), and In(5c) areas. In this adsorption configuration, the 6P molecule can readily avoid the O(3c)/In(5c), while interacting with as many In(5c) as possible. Thus, the variations in contrast in the AFM image, Figure 3(a), are the result of electronically undistorted phenyl rings that are situated on the O(3c) regions ("3" and "6"), and electronically modified or geometrically twisted phenyl rings due to the strong interaction with the In(5c) underneath ("2" and "5" equally, followed by "1"). Although the phenyl rings "4" and "1" are in a similar configuration the interaction with the CO functionalized tip is quite different (Figure 3(b)). This can be, for example, due to a strong geometric twist, or ring "1" and "4" differ electronically due to their different positions within the molecule. A variation in frequency shift along the molecule and particular on ring "4" is also observed in the constant-height AFM data acquired with metallic tips, see Figures 2(c) and 4(a). Apparently, 6P prefers under-coordinated Indium but tries to avoid under-coordinated O. Within the inhomogeneous and large unit cell of $In_2O_3(111)$, this is not entirely possible, and this is probably the reason why some phenyl rings twist from a flat orientation. This suggests that, by judiciously tuning atomic structure and the distribution of under-coordinated sites, one can "steer" an organic molecule to interact with a surface in different ways, presumably with interesting consequences for alignment of the frontier orbitals and charge injection. In summary, an essentially flat and planar adsorption geometry is proposed for single 6P molecules on $In_2O_3(111)$.

Increasing the sexiphenyl coverage results in a significant re-orientation of the molecules from the "dilute layer" into a (2×1) structure, once a critical coverage is reached. The molecules change from their preferred azimuthally oriented adsorption site of the single molecule into positions that are rotated +/-30° off of these directions (*i.e.*, from <1-10> to <-211>), and they adopt a densely packed structure. In STM this 6P layer shows a prominent dark-and-bright stripe pattern, where the individual molecules are difficult to discern with a metallic tip (Figures 5(b) and 6(c)). This contrast varies slightly for different tip preparations and shows almost no bias dependence in the empty states from +1 to +1.8 V. Sometimes the molecules within the dark stripes become more distinct (see Supporting information), featuring the zigzag "W" shape that is also characteristic of the single molecules aligned azimuthally in <1-10> orientations as discussed earlier (see Figure 2). This similarity suggests that both, the single molecules and those of the dark stripes are in the same, or at least a very similar, geometric configuration despite their different azimuthal orientations. The adsorption sites shown in Figure 6(b) were derived from intermediate coverages as in Figure 5(a). According to our evaluation, the molecules of the dark and bright stripes occupy non-equivalent sites on the surface. Those of the dark stripes occupy the site where the O(3c) (bound to In(5c)) are avoided most easily, while maximizing the contact to In(5c) (bound to O(4c)). The 6P molecules of the bright stripes have one end directly on the O(3c)/In(6c) region, and its site is mostly O(3c) terminated.

In the STM images, the molecules of the bright stripes appear somewhat shorter than those of the dark stripes. Moreover, the molecules within both the dark and the bright stripes are separated by 0.7 nm as measured perpendicular to their long axis (distance of the parallel black lines, indicated in blue in Figure 6(e)). The spacing of 0.7 nm also corresponds to the Van der Waals width of the flat 6P molecule (0.67 nm), suggesting a dense packing of the molecules if they indeed adsorb in an essentially planar fashion. Comparable values have been found for 6P on Cu(110) in the densely packed, planar and flat monolayer (molecular spacing of 0.72 nm) and for the twisted second layer there (molecular spacing of 0.67 nm).[19] An even closer arrangement occurs on O-passivated Cu(110)-(2×1) with 0.51 nm. This compression of the 6P(20-3) layer (bulk spacing of 0.566 nm and a tilt angle of ~33° with the 6P long axis as rotational axis) is realized by a slightly larger tilt of ~37°.[31,34] Also, on $TiO_2$(110), the flat-lying 6P tilts (rotation around its long axis) in the first monolayer grown at 130°C to accommodate the substrate lattice constant of 0.65 nm in [1-10] direction.[17]

The main reason for the azimuthal reorientation of the 6P molecules into the (2×1) first-monolayer structure is the molecule-molecule interaction that comes into play once the "dilute layer" cannot accommodate additional molecules due to steric hindrance. The formation of a first monolayer of 6P with molecules pointing in different directions is rather unusual. The bulk structures of 6P all feature molecules with parallel long axes; a herringbone arrangement is found only in the plane perpendicular to the long axis.[10,11] Thus the transition of 6P on $In_2O_3$(111) into the bulk phase either requires further restructuring of the (2×1) monolayer (probably starting with the (1×1) arrangement reported in this work) or it takes place in the subsequent layers.

The density of the "dilute layer" would be 0.5 6P molecules per substrate unit cell, *i.e.,* 0.5 ML, if a single domain were to ideally cover the whole surface. Experimentally, a higher coverage is observed due to the presence of three orientations that allow the molecules to share their substrate unit cells. The transition from the "dilute layer" into the (2×1) structure happens at coverages very close to 1 ML, compare Figures 1(c) and 5(a). The (2×1) unit cell contains two 6P molecules that corresponds to one 6P per $In_2O_3$(111) substrate unit cell, *i.e.,* to a molecular density of 1 ML = $5.64 \times 10^{13}$ cm$^{-2}$. Thus, the (2×1) monolayer doubles the number of molecules per unit cell with respect to the ideal "dilute layer".

In summary, we have reported on the adsorption of single molecules and the formation of the first layer of sexiphenyl on $In_2O_3$(111), as investigated with STM. For the single molecules, we find an unusual zigzag like appearance by STM and nc-AFM, suggesting a strong interaction with In(5c) atoms of the substrate, possibly distorting the molecule. The first monolayer features a (2×1) structure of flat-lying molecules with a density of one 6P molecule per substrate unit cell and molecules that are not uniaxially aligned.

*Supporting Information.* Deposition of 6P at room temperature; 6P on Cu(100) imaged with a CO tip; Appearance of the (2×1) structure in STM.

**Acknowledgements**
M.W. gratefully acknowledges the FWF project T759-N27. Research at the Oak Ridge National Laboratory for L.A.B. was sponsored by the U.S. Department of Energy, Basic Energy Sciences, Materials Sciences and Engineering Division (DE-AC05-00OR22725). M. Schmid acknowledges support from FWF project F4505. U.D. acknowledges support by the European Research Council (Advanced Grant "OxideSurfaces" ERC-2011-ADG_20110209).

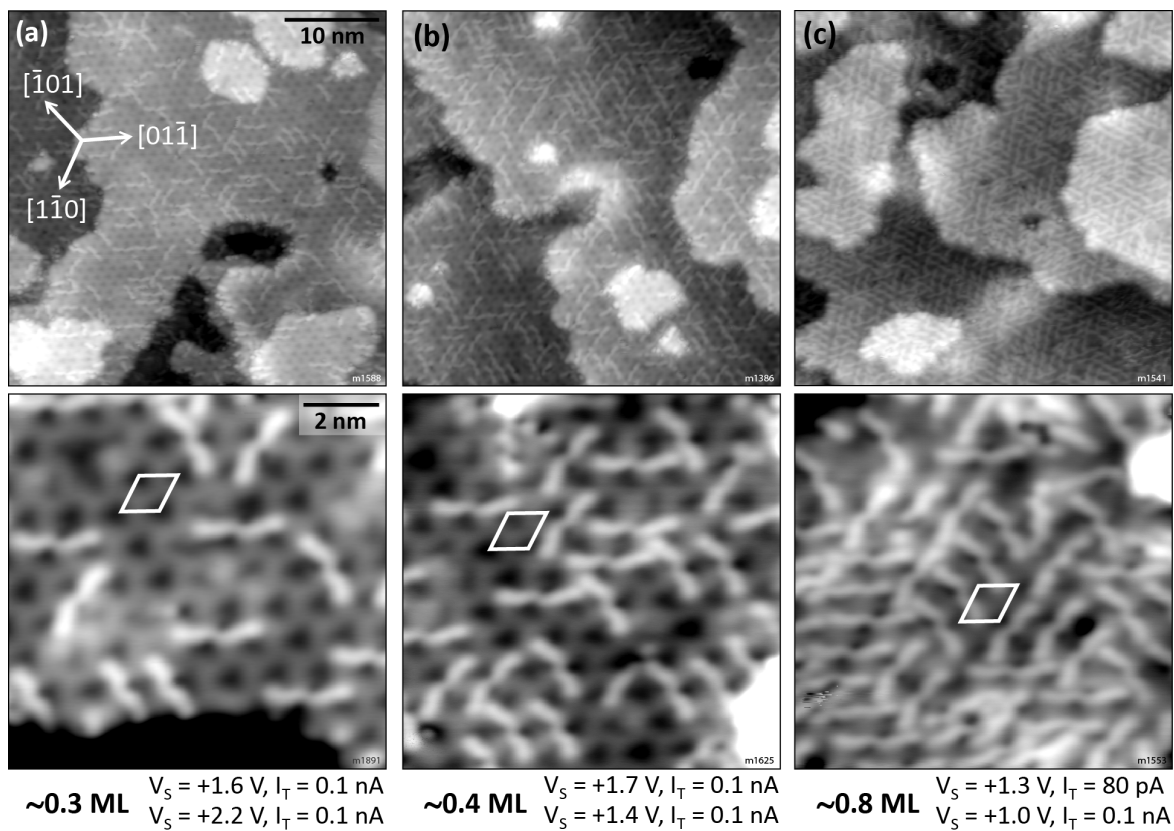

Figure 1: Low-coverage regime of sexiphenyl grown on $In_2O_3$(111) at 200 °C imaged with STM at 80 K. The unit cell of the $In_2O_3$ substrate is indicated in white. (a, b) Sub-monolayer coverages. (c) The dilute first layer.

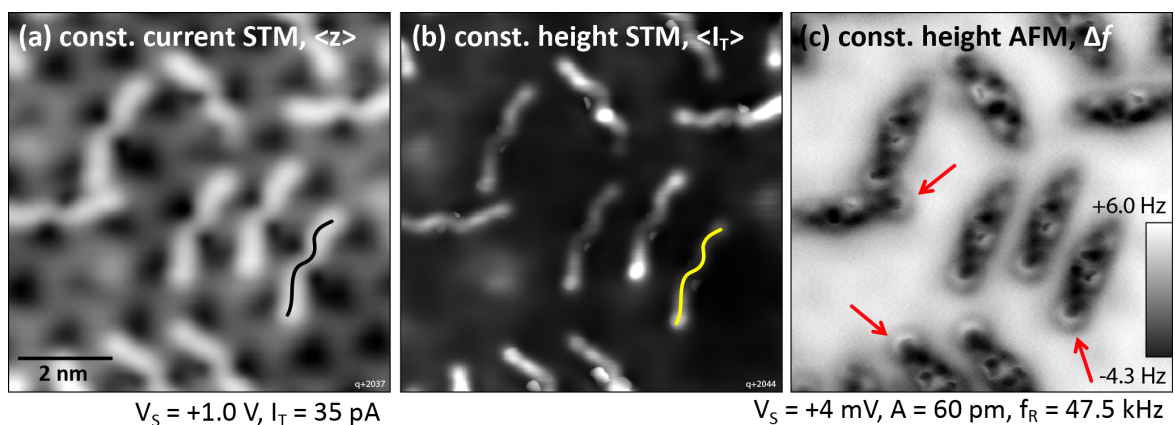

Figure 2: Single 6P molecules imaged with STM and nc-AFM at 5 K. (a, b) The molecules adopt an asymmetric "W" shape, as indicated in black/yellow. (c) Constant height nc-AFM (unknown tip) reveals an anisotropic pattern along the molecules' long axis. One end of each 6P (red arrows) shows a clearly discerned phenyl ring.

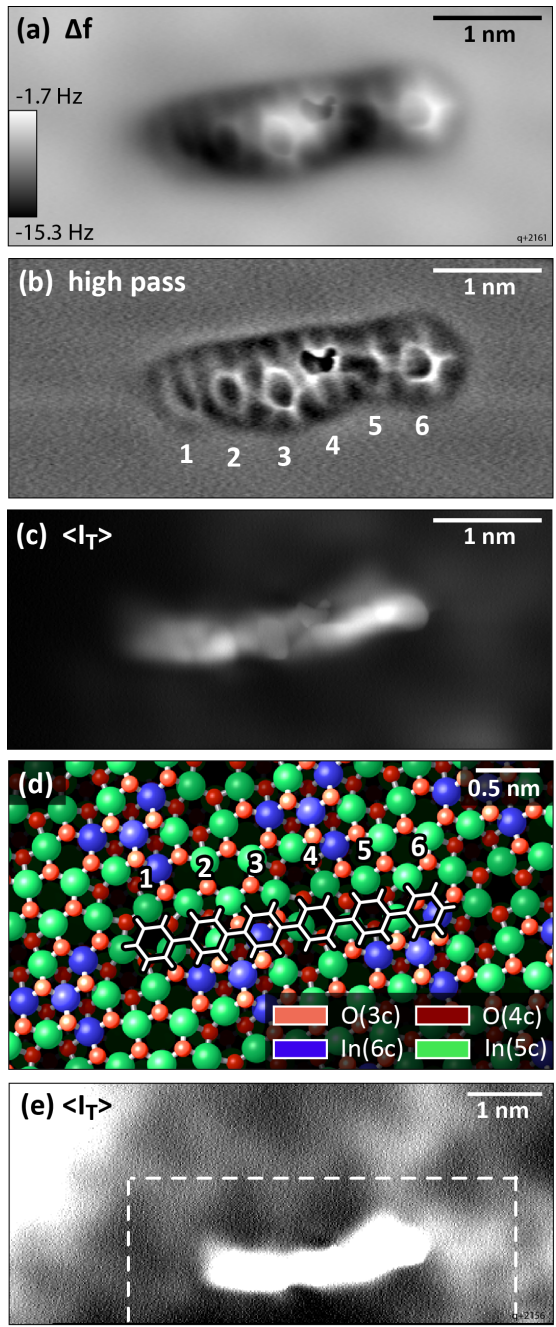

$V_S$ = +70 mV, A = 100 pm, $f_R$ = 74.5 kHz

Figure 3: Constant height nc-AFM images measured with a CO-terminated tip at 5 K. (a) Frequency-shift image. (b) High-pass filtered image of (a). (c) The simultaneously acquired tunneling current signal. (d) Simplified 6P structure superimposed on the atomic model of the $In_2O_3$(111) surface. (e) Constant-height STM image showing the substrate features used as reference for the adsorption site assignment in (d); the dark triangles in the STM images correspond to regions of the blue In(6c) atoms.

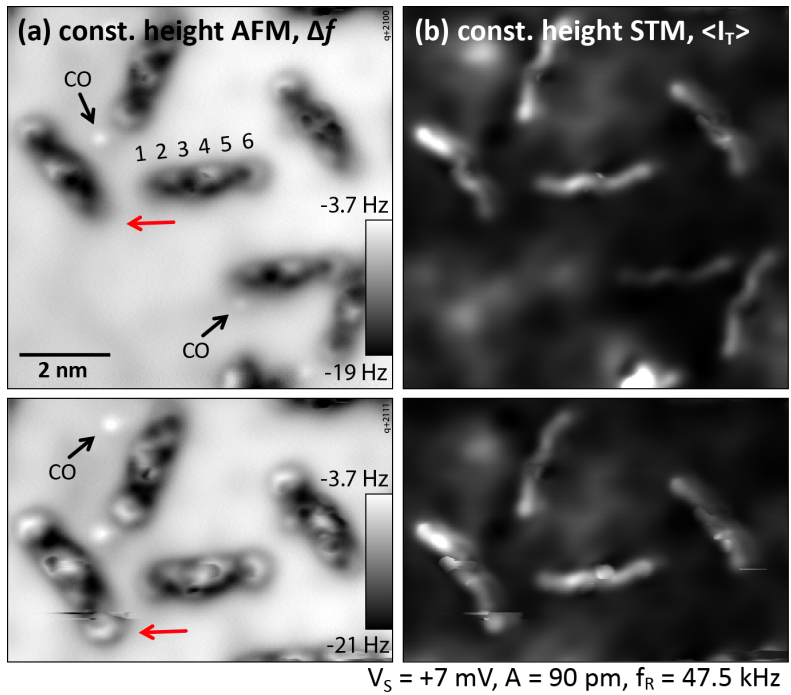

Figure 4: Identification of phenyl ring "1" (red arrow) by gently pushing it with the tip during the image acquisition. (a) Frequency shift signal. (b) Simultaneously recorded tunneling current. The slow scanning direction is from top to bottom. $T_{SPM}$ = 5 K.

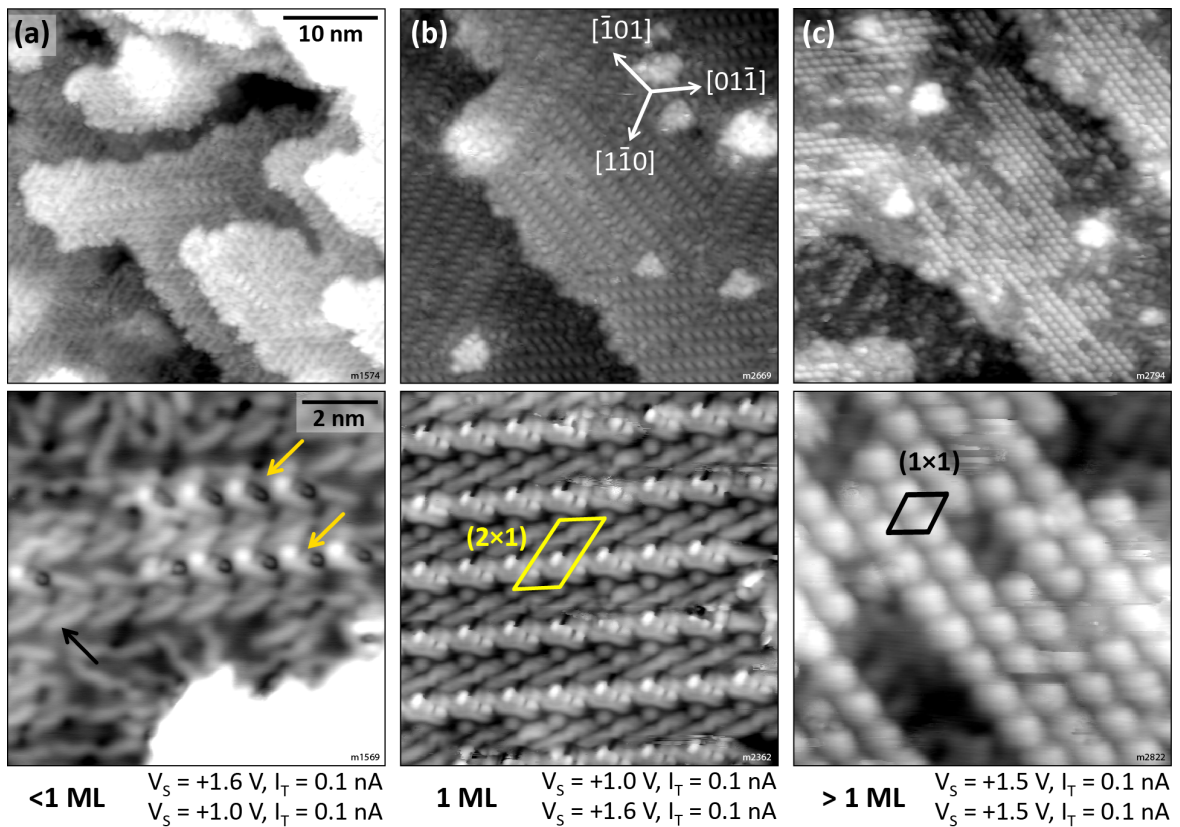

Figure 5: High-coverage regime of sexiphenyl grown on $In_2O_3$(111) at (a, b) 200°C and (c) 120°C. (a) Reorientation and onset of densely packed patches. (b) The densely packed (2×1) overlayer (defined as 1st monolayer). (c) Coverage beyond the densely packed layer with a (1×1) periodicity. $T_{STM}$ = 80 K.

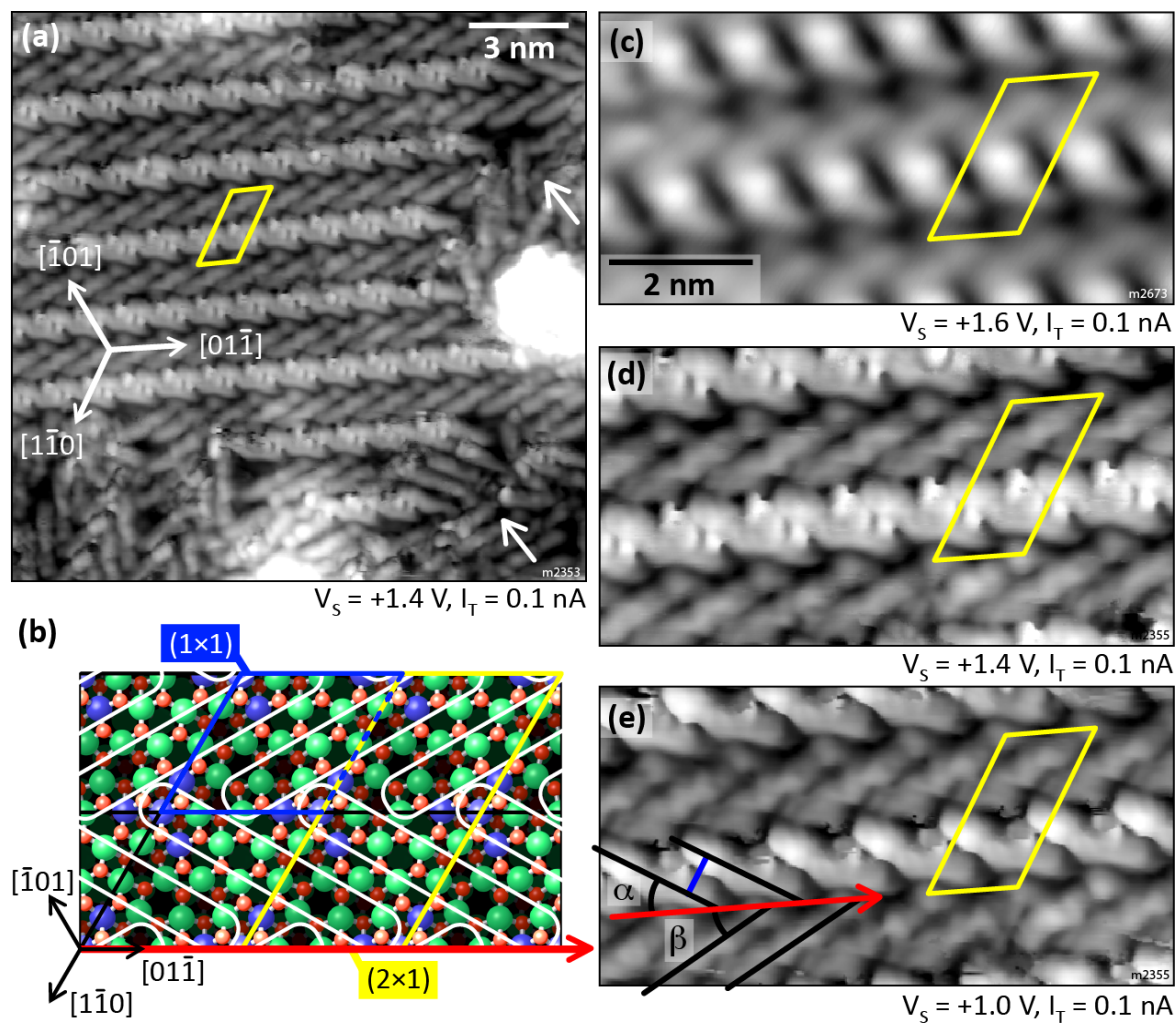

Figure 6: The densely packed (2×1) structure. (a) An ordered domain with loosely packed molecules at the fringes (white arrows). (b) Comparison of the $In_2O_3(111)$ (1×1) unit cell (blue) to the 6P (2×1) cell (yellow). The red arrow indicates the direction of the 6P stripes and the white ovals represent the molecules. The sites were determined from single stripes as in Figure 5(a). (c) Most common imaging contrast. (d, e) The same tip termination as in (a) at different bias voltages revealing the individual molecules. (e) The [01-1] surface direction is indicated by the red arrow to show the azimuthal orientation of the molecules (black lines) of $\alpha = \beta = 30°$. $T_{STM} = 80$ K.